\documentclass[journal]{vgtc}                     

\onlineid{1497}

\vgtccategory{Theoretical \& Empirical}

\title{Exposure to Common Data Visualization Types Among the U.S. Adult Population}

\author{%
  \authororcid{Kiegan Rice}{0000-0002-7454-7733},
  Nola du Toit, 
  \authororcid{Quentin Brummet}{0000-0003-3026-8983}, and 
  \authororcid{Heike Hofmann}{0000-0001-6216-5183}
}

\authorfooter{
  \item
  	Kiegan Rice is with NORC at the University of Chicago.
  	E-mail: rice-kiegan@norc.org.
  \item
  	Nola du Toit is with Research to Reach.
  	E-mail: dutoit.nola@gmail.com.
  \item 
    Quentin Brummet is with NORC at the University of Chicago.
  	E-mail: Brummet-Quentin@norc.org.
  \item
    Heike Hofmann is with the University of Nebraska-Lincoln.
    E-mail: hhofmann4@unl.edu.
}

\abstract{%
  Data visualizations are a primary means of communicating statistical information to the public. Their expanding use across news media, public health, education, and government reporting places greater importance on audiences’ ability to recognize and interpret them. While a significant body of prior research has established frameworks for the measurement of graph and visualization literacy, far less is known about everyday exposure to different kinds of charts and graphs among the general adult population. This gap limits our ability to meaningfully interpret differences in graph literacy, focus efforts on improving that literacy, and design visualizations that align with audience ability. Establishing population level estimates of exposure to data visualizations is therefore essential for improving visual communication and reducing misinterpretation of quantitative information. To address this gap, we surveyed a nationally representative sample of 1,168 U.S. adults via NORC’s AmeriSpeak panel about their exposure to sixteen different common data visualization types. The resulting survey responses demonstrate pronounced differences in baseline exposure to data visualizations across chart types and demographic groups. While widely used formats such as bar, pie, line, and grouped bar charts are highly recognizable to the vast majority of U.S. adults, many other practitioner-favored chart types remain unfamiliar to substantial portions of the adult population. Exposure also differs significantly by age and education, with higher educational attainment linked to greater exposure and older adults reporting lower exposure overall. Our findings provide essential population level context on visualization literacy and underscore the importance of aligning data visualization design with audience exposure and experience.
}


\teaser{
  \centering
  \includegraphics[width=\linewidth, alt={Sixteen simple icon images of different common data visualization types with their corresponding names.}]{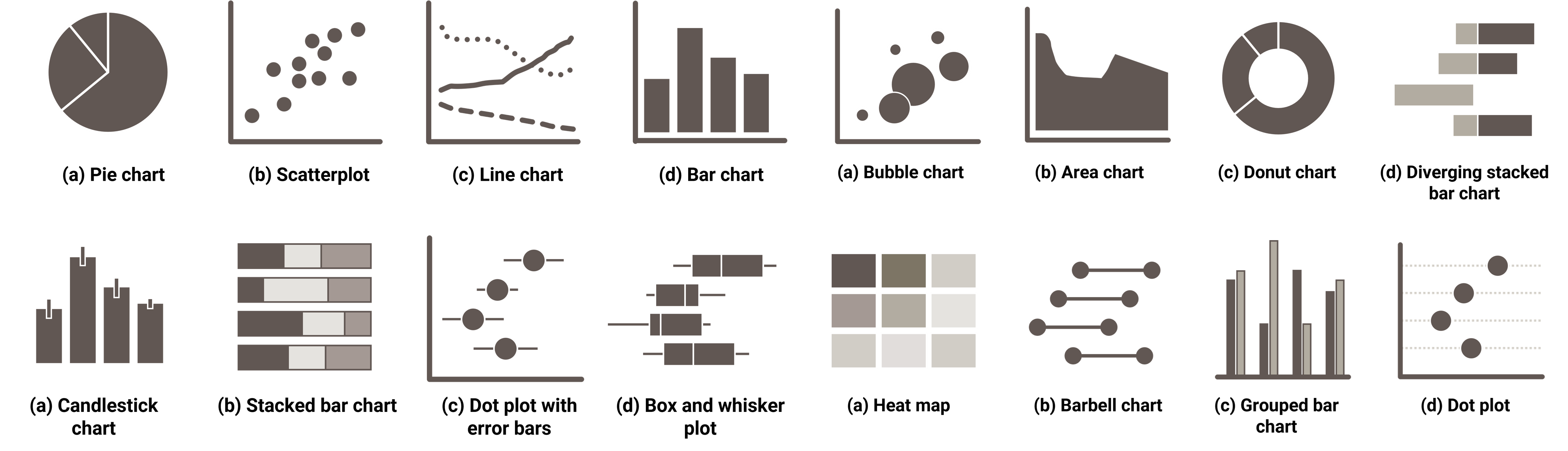}
  \caption{%
  	Iconographic representations of sixteen common data visualization types shown to study participants.%
  }
  \label{fig:teaser}
}

\graphicspath{{figs/}{figures/}{pictures/}{images/}{./}} 

\usepackage{tabu}                      
\usepackage{booktabs}                  
\usepackage{multirow}
\usepackage{lipsum}                    
\usepackage{mwe}                       
\usepackage{ccicons}                   
\usepackage{xcolor}                    

\usepackage{mathptmx}                  

\begin{document}


\firstsection{Introduction}

\maketitle

Data visualization plays a central role in how scientific information is communicated to non-expert audiences. In contemporary public communication, data appears throughout news media, dashboards, health portals, educational reports, and government websites; visualizations have become a primary interface through which the public encounters scientific, policy, and risk information. Although some individuals learn to read and interpret visualizations through formal education, many now encounter them incidentally in everyday life, outside structured instruction. As a result, informal exposure increasingly complements formal learning in shaping how people understand common graphical formats. Understanding how these different exposure pathways influence comprehension is essential for connecting visualization design practices with how the public actually interprets them.

Prior work on graph familiarity shows that people reason differently about statistical information when the visualization format is familiar. Existing experience with graphics can improve the ability to interpret and interact with them \cite{hallProfessionalDifferencesComparative2022, tandon_measuring_2022}. Familiar graphics are often seen as easier to understand, particularly by non-experts \cite{quispelGraphChartAesthetics2016}. This motivates an important, yet understudied question: \emph{How much exposure to data visualizations do individuals have?}

This question becomes especially important because the large literature on visual literacy often focuses on small sets of relatively simple and common chart types, such as bars and lines, where it may be assumed that most study participants already have a passing familiarity with the chart type. In studies that cover a broader range of chart types or domain-specific visuals, it is difficult to untangle an individual's ability from familiarity; when literacy appears low, it may reflect a lack of exposure rather than a general inability to read charts. Understanding even basic differences in exposure is therefore crucial. For this question in particular, evidence from a representative sample is essential to examine how familiarity varies across individuals from different demographics.

To address this gap, we measure the population’s exposure to a set of common data visualization types. Using a probability-based sample of more than 1,000 adults aged 18 and older, we ask respondents about their exposure to 16 chart types. This approach provides estimates of overall exposure and leverages the sample’s representativeness to document how that exposure differs across demographic and experiential subgroups.

The results offer the first nationally representative estimates of exposure to widely used graphics. This creates a launching point for examining how literacy might vary for more complex chart types and across individuals. The results also establish a foundation for designing studies of graphical understanding and for tailoring visualization design to different audiences.

\section{Background}
Data visualization literacy, or graph literacy, has been defined in several ways, and broadly involves a viewer's ability to read, understand, and interpret information presented in data visualizations. There is a rich history of studying graph literacy; prior studies range from lab experiments to classroom studies to crowd-sourced web-based user studies to visualization tests "in the wild" in more public settings. These studies focus on various elements of visualization literacy, including testing basic ability to identify data points in a chart, make value judgments or comparisons, describe trends, assess or make real-world interpretations from a chart, make risk assessments based on data, identify incorrect or misleading charts, and judge how personally useful a chart is for a viewer. These disparate tasks emphasize that visualizations have different goals, audiences, and outcomes and that visualization literacy measurement is a complex construct that includes multiple levels of understanding and usage. The aforementioned increase in informal data visualization exposure makes this literature even more critical, as visualizations are increasingly used to guide decision-making.  

Different models have been proposed to measure visualization literacy/graph literacy at these distinct levels of understanding. Although terminology in the field is not always consistent \cite{bornerDataVisualizationLiteracy2019}, several themes recur across the literature. One common framework is Bloom's taxonomy, which is frequently used to conceptualize levels of learning applied to visual comprehension \cite{bloomTaxonomyLearning}. For instance, Burns et al. developed test items to measure visualization literacy directly based on the levels of the taxonomy, ranging from tests of basic knowledge (retrieve points; locate values; identify axis labels) to ability to evaluate (justify conclusions based on data; judge which design is more appropriate)  \cite{burns_how_2020}. Such work helps shape the development of visualization literacy frameworks and empirical studies, which vary widely in their targeted levels of understanding and approaches to assessment.

\begin{table*}[!hbt]
\centering
\scriptsize
\resizebox{\textwidth}{!}{
\begin{tabular}{lccccccccccc}
\toprule
 & \multicolumn{11}{c}{\textbf{Chart Type}} \\
\cmidrule(lr){2-12}
\textbf{Study}
& Bar
& Line
& Pie
& Stacked Bar
& Choropleth Map
& Scatterplot
& Grouped Bar
& Table
& Bubble Chart
& Treemap
& Icon Array \\
\midrule
Wainer (1980) 
& X & X & X &   &   &   &   & X &   &   &   \\
Galesic \& Garcia-Retamero (2011)
& X & X & X &   &   &   &   &   &   &   & X \\
Boy et al. (2014) 
& X & X &   &   &   & X &   &   &   &   &   \\
Nayak et al. (2016)
& X & X &   &   &   &   &   & X &   &   &   \\
Börner et al. (2016)
&   & X & X & X & X &   & X &   & X & X &   \\
Lee et al. (2017) -- \emph{VLAT}
& X & X & X & X & X & X &   &   & X & X &   \\
Peck et al. (2019)
& X & X &   &   & X &   & X &   &   &   &   \\
Durand et al. (2020)
& X &   &   &   &   &   &   & X &   &   & X \\
Burns et al. (2020)
& X & X &   & X &   &   & X &   &   &   &   \\
Hall et al. (2022)
&   &   & X &   &   & X &   &   &   &   &   \\
Tandon et al. (2023)
& X &   &   &   &   &   &   &   &   &   &   \\
Pandey \& Ottley (2023) -- \emph{mini-VLAT}
& X & X & X & X & X & X &   &   & X & X &   \\
Ge et al. (2023) -- \emph{CALVI}
& X & X & X & X & X & X &   &   &  &  &   \\
Rice et al. (2026) 
&  & X &  & X &  &  &   &   &  &  &   \\
Saske et al. (2026) -- \emph{MdamV}
& X & X &  &  &  &  &   &   &  &  &   \\
\midrule
Current study
& X & X & X & X &  & X &  X &   & X &  &   \\
\bottomrule

\end{tabular}
}
\caption{Chart types used across visualization literacy and graphicacy studies. Columns include only chart types appearing in two or more studies and are ordered by frequency of inclusion. Chart types appearing in only one study are omitted from the table and include: area chart, heat map, histogram, infographic, isotype, radial bar chart, stacked area chart, word cloud, Sankey diagram, network diagram, and isocontour plot. Note that Lee et al. (2017), Pandey and Ottley (2023), and Ge et al. (2023) distinguish between stacked bar and 100\% stacked bar charts and both are included in their assessment tests; here, we group them together into the broader stacked bar category. Similarly, Rice et al. (2026) distinguish between stacked bar charts and diverging stacked bar charts, which are grouped into stacked bar charts here for simplicity.}
\label{tab:chart-types-literature}
\end{table*}

Several formal task-based tests measure individuals' visualization literacy have emerged over the years. Early work by Wainer tested graphicacy among children using simple bar charts, line charts, pie charts, and tables \cite{wainer_test_1980}. Later, Galesic and Garcia-Retamero developed a graph literacy scale using simple bar charts, pie charts, line charts, and icon arrays \cite{galesic_graph_2011}; Boy et al. developed test items on bar charts, line charts, and scatterplots \cite{boyPrincipledWayAssessing2014}. A more comprehensive approach was proposed in Lee et al.'s Visualization Literacy Assessment Test (VLAT), which measures various levels of viewer comprehension across twelve distinct chart images using a 53-item test covering eight potential data visualization tasks, ranging from straightforward value retrieval tasks to more complex assessments of correlations or trends in the data, and including tasks of multiple difficulty levels for each chart type \cite{lee_vlat_2017}. A slightly modified 41-item version was then used to compare visualization literacy to numeracy and other cognition characteristics \cite{leeCorrelationUsersCognitive2019}. Pandey and Ottley developed a shortened version of this assessment--the “mini-VLAT”--which simplified the assessment to one question from each of the twelve chart types used in VLAT \cite{pandey_minivlat_2023}. Building on the VLAT framework, Ge et al. designed the CALVI assessment for evaluating critical thinking in visualizations. CALVI includes nine of the original VLAT chart types and further incorporates “trick” questions \cite{ge_calvi_2023}. Cui et al. subsequently proposed adaptive testing versions of both VLAT and CALVI \cite{cuiAdaptiveAssessmentVisualization2024}. Most recently, Saske et al. proposed the Multidimensional Assessment Method for Visualization Understanding (MdamV), an approach designed to integrate task-based ability, self-reported ability, and open-ended critique \cite{saskeMultidimensionalAssessmentMethod2026}. Note that across this literature there is substantial variability in measurement approaches, a key challenge called out by Ge et al. for the field of visualization literacy \cite{geAutoethnographyVisualizationLiteracy2026}. Even among four of these studies anchored in the same VLAT framework, there are notable differences in their goals and the levels of understanding they measure across chart types.

\cref{tab:chart-types-literature} provides a summary of chart types included in recent visualization literacy and graphicacy studies. With the exception of bar, line, and pie charts, there is not consistent coverage of visual formats across studies; while some recent studies include bar chart variations (i.e., stacked bars and grouped bars), they are not ubiquitous in visualization literacy tests or assessments. Chloropleth maps, scatterplots, bubble charts, tables, and treemaps have been utilized in multiple studies, though notably some of the repetition occurs due to mini-VLAT \cite{pandey_minivlat_2023} directly replicating the charts used in VLAT \cite{lee_vlat_2017}. Icon arrays were also used in multiple studies, primarily in the health communication literature where use of icon imagery is relatively more common. In other domain-specific studies, such as Durand et al.'s study of museum visitors, other information visualization formats have been used such as network diagrams and Sankey diagrams. Chart formats that include both values and a measure of uncertainty (e.g., box and whisker plots, dot plots with error bars) have been tested very little despite a growing literature on the difficulties of visually communicating uncertainty \cite{hullman_hypothetical_2015, hullman_imagining_2018, wittImpactFamiliarityVisualizations2021, franconeri_science_2021}. Several commonly tested chart types also have visual alternatives that have grown in popularity among data visualization practitioners in recent years, such as the use of donut charts as a pie chart alternative, dot plots as a bar chart alternative, and barbell charts as a grouped or stacked bar chart alternative. Despite their popularity among practitioners, it is unclear how familiar these alternative chart types are among general audiences or how viewer comprehension may or may not differ from standard bar and pie chart comprehension. 

Beyond variability in chart types, different studies in the literature study widely varying populations. Several studies have focused on specific sub-populations, with the aim of understanding how a target audience for visuals experiences and comprehends them given that specific audiences' visual literacy may differ from the highly-educated populations often used in visualization studies. For example, Börner et al. tested visualization literacy among a sample of 273 museum visitors, focusing on complex displays of graphical information that may support museum patrons: charts, maps, complex graphs, and network diagrams \cite{bornerInvestigatingAspectsData2016}. Peck et al. studied a hard-to-reach population, surveying 42 adults in rural Pennsylvania about 10 data graphics, including two infographics, and found that how viewers understand and utilize charts can differ based on their personal life experiences \cite{peckDataPersonalAttitudes2019}.
While these targeted studies provide valuable insights, a key goal of being able to effectively measure visualization literacy is to understand visualization literacy writ large in the population. To this end, Galesic and Garcia-Retamero produced population-level estimation studies of both statistical numeracy and visual literacy, with the latter focusing on bar charts, line charts, pie charts, and icon arrays and demonstrating that over one-third of U.S. adults have low graph literacy and low numeracy skills \cite{galesic_statistical_2010, galesic_graph_2011}. Rice et al. also demonstrated key differences in ability to correctly interpret information in stacked bar charts, diverging stacked bar charts, and line charts across educational attainment groups in the U.S., and further identified chart-specific differences in interpretation ability \cite{riceMeasuringRealWorldUnderstanding2025}.

The variability across sub-populations and, in particular, low visual literacy among specific groups, has tangible real-world consequences on how well important information is understood. For example, in patient-centered health communication research, patients with lower graph literacy on bar charts, line charts, and tables have exhibited greater difficulty interpreting graphical summaries of health information, even when their general health literacy and level of education are high \cite{nayakRelevanceGraphLiteracy2016}. This dynamic is even more concerning for vulnerable populations; graph literacy was shown to be lower in Medicaid-eligible populations than among the general U.S. population \cite{durand_graph_2020}. 

This points to two key gaps in the prior literature. First, there is limited understanding of how familiar the general population is with common chart types. Because studies of visual literacy use inconsistent sets of visualizations, familiarity becomes an important lens for interpreting their findings: what appears to be low literacy may instead reflect limited exposure or familiarity rather than a lack of interpretive ability. This may be particularly true when trying to compare relatively common chart types such as pie and bar charts with less common chart types such as dot or barbell plots. Solen et al. argued that visualization literacy should move towards measurement of a more nuanced overall visualization skillset, including distinguishing ability between familiar and unfamiliar charts \cite{solenVisualizationLiteracySkillset2025}.

In addition, many of the referenced visualization literacy studies have utilized small and/or convenience samples to study literacy. In a systematic review of data visualization literacy studies, Beschi et al. identified a median sample size of 86 participants and found that over 50\% of studies utilized crowd-sourcing platforms \cite{beschiCharacterizingDataVisualization2025}.  Studies that \emph{have} focused on population-level estimation of visualization literacy have been implemented with a narrow set of the most common data visualization formats; a review by Firat et al. identified bar charts, pie charts, line charts, and scatterplots as the most commonly tested chart types across visualization literacy studies, including classroom-based studies \cite{firatInteractiveVisualizationLiteracy2022}. This means that although existing assessments allow us to measure differences in visualization literacy for the most popular chart types, we still lack a clear understanding of literacy levels in the general population for many newer or more specialized data visualization formats.

Our study helps fill these gaps by focusing on measuring exposure across a number of chart types, including both commonly used ones and more modern practitioner-favored alternatives. Further, our use of a large, nationally representative sample of U.S. adults provides a comprehensive understanding of exposure overall and across groups, which identifies actionable insights and guidance on both design of data visualizations for public data communication as well as a framework for better understanding the "why" behind demonstrated literacy gaps.

\section{Study Design}

We selected sixteen visualization types to test, including common data visualization formats, modern practitioner-favored alternatives, and versions incorporating uncertainty into the visuals. The common visualization formats were selected to cover the most frequently-used visualization formats used in visualization literacy studies and included bar charts, line charts, pie charts, stacked bar charts, scatterplots, grouped bar charts, and bubble charts. Chloropleth maps were not included as they are distinct from many of the other common formats in representing spatial information. For practitioner-favored alternatives, we use donut charts, barbell charts, dot plots, and diverging stacked bar charts. Diverging stacked bar charts are a favorite among practitioners communicating survey or opinion research results, while donut charts are a common alternative to pie charts, barbell charts are an emerging replacement for grouped bar charts, and dot plots are a common replacement for traditional bar charts. Formats incorporating uncertainty representation encompass candlestick charts, dot plots with error bars, and box and whisker plots. We limited our study to sixteen chart types to mitigate respondent burden on answering a large number of items.

\begin{figure}[!b]
  \centering
  \includegraphics[width=\columnwidth, alt={Example of visual interface for a set of four charts shown to participants as part of the survey. Participants saw a series of four screens, with four chart types each.}]{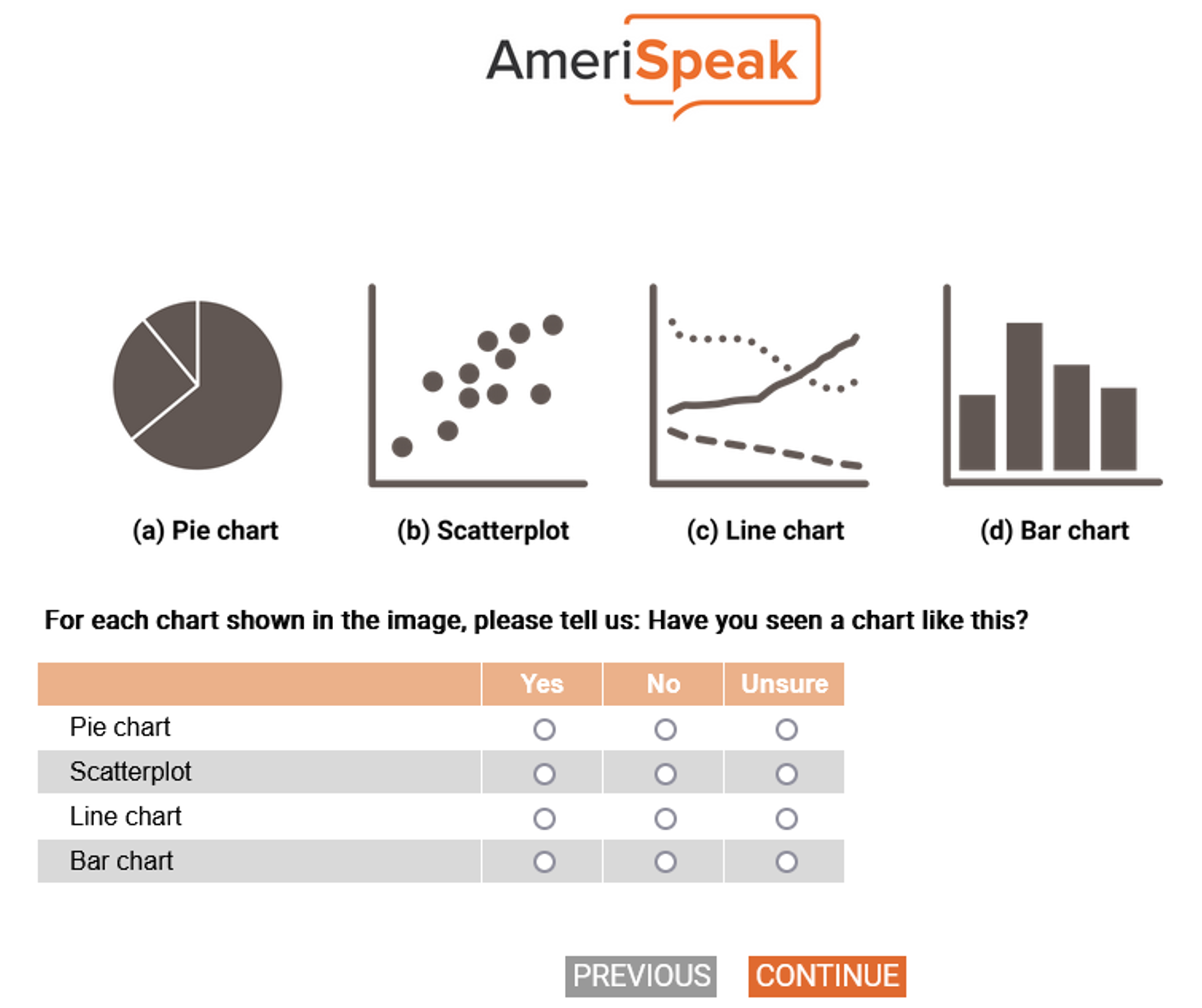}
  \caption{%
    Example of visual interface for a set of four charts shown to participants as part of the survey. Participants saw a series of four screens, with four chart types each. 
  }
  \label{fig:example_survey_view}
\end{figure}

Survey participants were shown a series of sixteen items, grouped on four distinct survey screens. Each screen contained a combined image of four iconographic representations of data visualization types and their corresponding names, labeled (a) through (d). On the corresponding screen, participants were then asked 'Have you seen a chart like this?' and given response options of 'Yes', 'No', and 'Unsure' for each chart type shown. An example screen for the first four items is shown in \cref{fig:example_survey_view}. Items were grouped into sets of four to reduce the respondent burden of clicking through sixteen individual screens while making each displayed set small enough so users could easily pair each icon and image with the corresponding question on-screen. Iconographic images were used to represent simplified versions of visual formats without requiring the participant to recognize a specific design element or type of data. Exposure to bar charts, for example, may encompass a wide variety of presentations, designs, data types, and contexts and two individuals reporting exposure to bar charts may have vastly different connections to the idea of a bar chart. Chart names were provided with the icons to mitigate the ambiguity introduced by that simplified form. Participants could thus be reporting exposure to the named chart, the iconographic representation, or both. The full set of all sixteen iconographic images and chart names shown to users are displayed in \cref{fig:teaser}.

Note that some chart types are commonly referred to by more than one name. In particular, a box and whisker plot is often called a boxplot, a barbell chart may also be known as a dumbbell plot, and a candlestick chart as shown in our iconographic image might also described as a dynamite plot by some audiences. For this reason, providing iconographic images in addition to chart names was essential to ensure respondents could reliably differentiate among chart types despite overlapping terminology.

\section{Study Participants}

Data were collected using NORC's AmeriSpeak Omnibus, a biweekly survey administered to a probability-based sample of respondents from a standing panel of more than 54,000 individuals aged 13 and older\cite{AmeriSpeakTechOverview2024}. The panel is drawn from U.S. households using NORC's National Sample Frame, which covers over 97\% of households nationwide. Each wave of the Omnibus survey yields completed surveys from approximately 1,000 respondents aged 18 and over. All participants complete a consent statement as part of their participation in the AmeriSpeak panel.

\begin{table*}[hbt!]
\centering
\begin{tabular}{llrrr}
\hline
Demographic & Group & N Respondents & Weighted \% of Respondents (s.e.) & Population Totals \\
\hline
\multirow{5}{*}{Education Level}
 & Less than HS & 66 & 9.0 (1.30) & 8.4 \\
 & HS graduate or equivalent & 210 & 28.8 (1.88) & 28.7 \\
 & Some college/associates degree & 458 & 26.4 (1.45) & 26.3\\
 & Bachelor's degree & 239 & 20.0 (1.47) & 22.8 \\
 & Post grad study/professional degree & 195 & 15.8 (1.29) & 13.8 \\
\hline
\multirow{4}{*}{Income Level}
 & Less than \$30,000 & 222 & 21.8 (1.69) & 14.6\\
 & \$30,000 to under \$60,000 & 313 & 26.1 (1.62) & 22.1 \\
 & \$60,000 to under \$100,000 & 275 & 21.6 (1.48) & 23.7 \\
 & \$100,000 or more & 358 & 30.5 (1.73) & 39.7 \\
\hline
\multirow{4}{*}{Age Group}
 & 18--29 & 188 & 19.6 (1.60) & 19.7 \\
 & 30--44 & 364 & 25.5 (1.53) & 25.9 \\
 & 45--59 & 284 & 23.8 (1.59) & 23.1 \\
 & 60+ & 337 & 31.1 (1.79) & 31.3 \\
\hline
\multirow{2}{*}{Metro Area Residency}
 & Non-Metro Area & 175 & 13.5 (1.24) & 13.3 \\
 & Metro Area & 993 & 86.5 (1.24) & 86.6 \\
\hline
\multirow{2}{*}{Gender}
 & Male & 570 & 48.6 (1.89) & 48.7 \\
 & Female & 598 & 51.4 (1.89) & 51.3 \\
\hline
\multirow{4}{*}{Race/Ethnicity}
 & White, non-Hispanic & 722 & 60.5 (1.87) & 60.5 \\
 & Hispanic & 202 & 18.0 (1.52) & 18.0 \\
 & Black, non-Hispanic & 152 & 12.1 (1.19) & 12.2 \\
 & Asian-Pacific Islander, non-Hispanic & 51 & 6.1 (1.00) & 6.9 \\
 & Other, non-Hispanic & 41 & 3.2 (0.64) &  2.4 \\
\hline
\end{tabular}
\caption{Distribution of survey participants by self-identified education level, income level, age group, metropolitan area residency, gender, and 4-category race/ethnicity group. The N respondents column shows the number of individual participants self-identifying in each group who completed the survey. The weighted percent of respondents column shows the percent of participants self-identifying in each group and corresponding standard error after applying survey weights. Population totals are sourced from Current Population Survey (CPS) data for October 2024, when the survey was fielded.}
\label{tab:participant_summary}
\end{table*}

Probability-based sampling was conducted across 48 strata defined by age, education, gender, and race and Hispanic ethnicity. Stratum sizes were aligned with corresponding population distributions and adjusted for anticipated differential response rates to support the generation of a representative sample of U.S. adults. Final survey weights were calibrated to benchmarks from the U.S. Census Bureau's Current Population Survey (CPS) and balanced across gender, age, educational attainment, race and ethnicity, and geographic region. 

Given the visual nature of this study, our survey items were administered only via web, and respondents who took the survey by telephone were not asked these questions. Web-mode participants constitute 93.3\% of the panel overall and account for 96.3\% of the panel's weighted representation.  

Data were collected as part of an Omnibus wave in October 2024, with fielding occurring from October 24 - 28, 2024. A total of 1,173 responses were gathered; five participants were removed due to skipping all sixteen items, resulting in a total sample size of 1,168. A small subset of participants (n = 61, 5.2\% of total completes) skipped one or more individual items among the whole set of sixteen, and their responses are included in the analyses below when non-missing. The item-level non-response rates were relatively low, and only 3 of those 61 respondents skipped 5 or more items. The raw sample sizes, weighted percentage by demographic groups, and population totals from the CPS are shown in \cref{tab:participant_summary}. These results show alignment between the sample in the study and the CPS, supporting the representativeness of the sample.

\section{Results}

We investigate the resulting survey responses using both descriptive statistics and formal statistical modeling to measure differences across groups and chart types. Weighted survey files were used throughout analysis, and survey weights were applied in all statistical modeling and when calculating variable means and weighted percentages of responses. Analyses were completed in R primarily using the \texttt{survey} package for survey data analysis \cite{surveyPackage}. Survey responses and code to produce all results are provided as supplemental materials.  

\subsection{Exposure Across Types}

\begin{figure}[!b]
  \centering 
  \includegraphics[width=\columnwidth, alt={A multi-panel bar chart showing the distribution of response types. The top row visualizes the number of times participants said 'Yes', that they had seen a chart before. The middle row visualizes the number of times participants said that they were 'Unsure'. The bottom row depicts the number of times participants said 'No', that they had not seen the chart before.}]{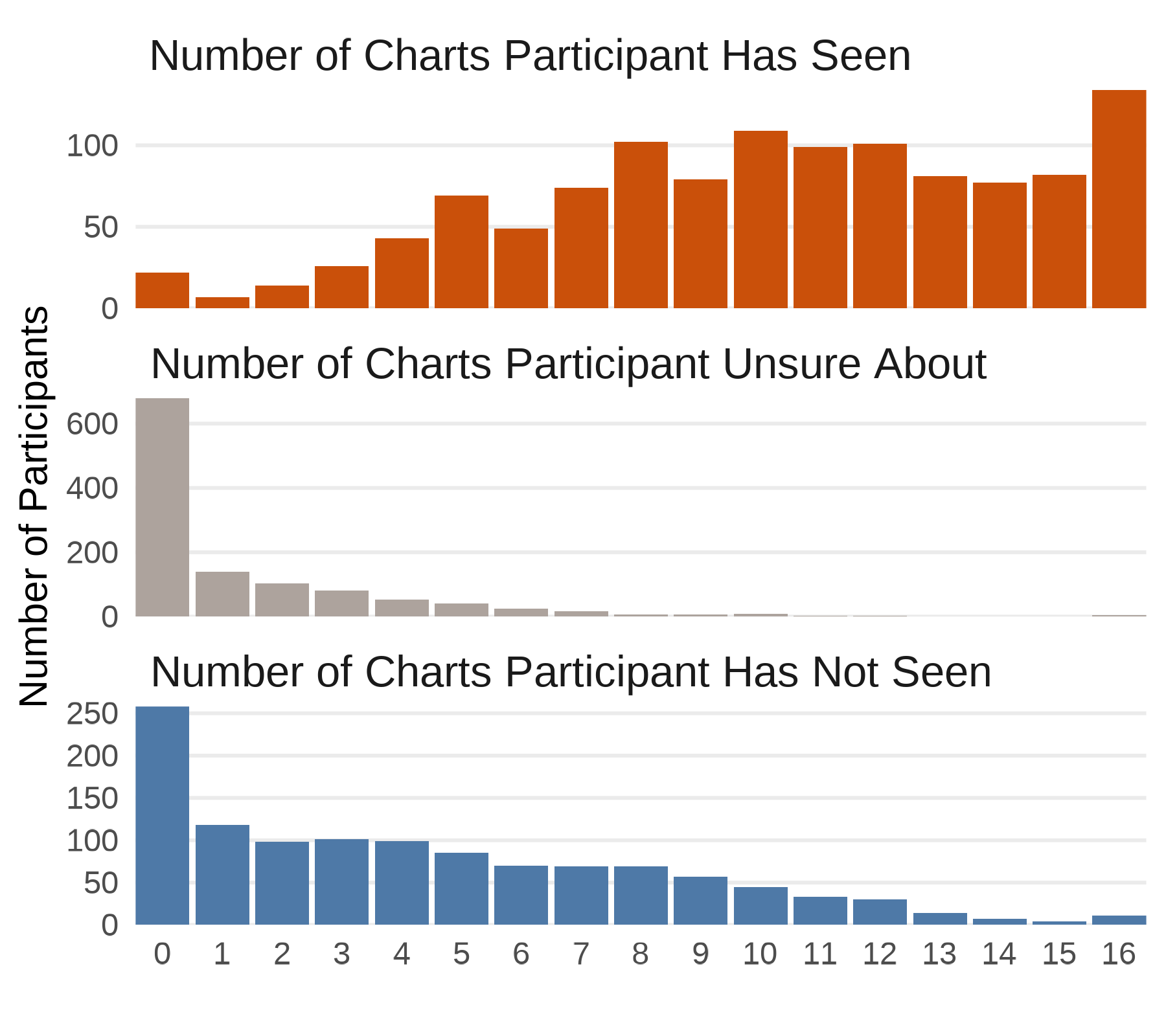}
  \caption{%
  	Visualization of the number of Yes, No and Unsure responses for participants across all sixteen items.%
  }
  \label{fig:response_distributions}
\end{figure}

\cref{fig:response_distributions} shows the overall distribution of exposure responses across each respondent's sixteen responses. Both number of Unsure and No responses are right skewed, with over half of respondents (58.3\%) reporting 'Unsure' on zero out of sixteen items and nearly a quarter (22.4\%) of respondents reporting 'No' on zero out of sixteen items. 

Although more than 40\% of respondents were unsure about at least one chart type, a very small proportion (2.2\%, 26 total respondents) reported being unsure about more than half (9 or more) of the chart types shown. Respondents had relatively high certainty as to whether they had or had not seen most of the chart types displayed. 

11.4\% of respondents reported 'Yes' to all sixteen items, meaning they reported having seen every single chart type shown to them. The median number of familiar chart types across all respondents was 10, and the weighted survey mean was 9.87 (SE = 0.16).

To examine the distribution of exposure to chart types across groups, we fit a logistic regression model for the proportion of 'Yes' responses out of 16 total items for respondent $k$: 

\begin{equation}
\label{eq:logistic_model_overall}
\mbox{logit}\ P(Y_{k}) = \mu + X_k\beta
\end{equation}

where $X_k$ represents a vector of demographic characteristics for participant $k$, including 5-category education level, 4-category income level, 4-category age group, binary metropolitan area residency, gender, and 5-category Race/Ethnicity. The resulting model coefficient estimates and average marginal effects are displayed in \cref{tab:ame_demographics_model}. At a high level, the results indicate that education and age are the strongest predictors of exposure with data graphics, with smaller associations observed for income, gender, and race.

Turning to specific results in \cref{tab:ame_demographics_model}, respondents with a high school diploma or equivalent do not differ statistically in their exposure from those with less than a high school education. However, exposure increases substantially among those with postsecondary education. Individuals with some college or an associate degree are report exposure to 18 percentage points more of the graphics on average, conditional on other covariates. Reported exposure continues to rise with additional education, and respondents with postgraduate study or a professional degree exhibit the highest levels of exposure, increasing up to a 28 percentage point gain in the percentage of recognized graphics.

Age is the next strongest predictor. Exposure declines consistently with age, with each successive age group reporting lower exposure than the one before it. Respondents aged 60 and older have, on average, been exposed to 22 percentage points fewer of the graphics than younger adults.

Income is more modestly associated with exposure, and none of the income groups below \$100,000 differ significantly from one another. Respondents with incomes above \$100,000, however, report exposure to approximately 7 percentage points more of the graphics than those with incomes under \$30,000, conditional on other covariates.

Residence in metropolitan versus non-metropolitan areas is not significantly associated with exposure after adjusting for other covariates.

Finally, gender and race exhibit relatively modest associations. Women report exposure to about 5 percentage points fewer of the graphics than men; although statistically significant, this difference is smaller than the effects for age and education. Estimates also suggest that non-Hispanic Black respondents are less likely than non-Hispanic White respondents to report exposure to the graphics, and the magnitude of this difference is similar to male-female differences.

\begin{table}[!bt]
\centering
\caption{Regression coefficients and average marginal effects for the logistic regression model estimating proportion of 'Yes' responses, as defined in \cref{eq:logistic_model_overall}. Significance of model coefficients denoted as follows: $^{*} p<.05$; $^{**} p<.01$;  $^{***}p<.001$. Average marginal effects are significantly different from zero if the 95\% confidence interval does not include zero. Reference groups that are part of the model intercept shown with dashes.}
\label{tab:ame_demographics_model}
\resizebox{\columnwidth}{!}{%
\begin{tabular}{lrlr}
\toprule
 & Estimate (S.E.) & $p$-value & AME (95\% CI) \\
\midrule
Intercept \\
\quad Intercept
& 0.44 (0.20) & 0.031* & NA \\
\midrule

Education level \\
\quad Less than HS
& -- & -- & -- \\
\quad HS graduate or equivalent
& 0.26 (0.19) & 0.171 & 0.06 (-0.03, 0.15) \\
\quad Some college/associates degree
& 0.78 (0.18) & <.001*** & 0.18 (0.10, 0.27) \\
\quad Bachelor's degree
& 1.00 (0.19) & <.001*** & 0.23 (0.14, 0.32) \\
\quad Post grad study/professional degree
& 1.25 (0.20) & <.001*** & 0.28 (0.19, 0.37) \\
\midrule

Income level \\
\quad Under \$30{,}000 
& -- & -- & -- \\
\quad \$30{,}000 to under \$60{,}000
& -0.09 (0.12) & 0.455 & -0.02 (-0.07, 0.03) \\
\quad \$60{,}000 to under \$100{,}000
& 0.11 (0.13) & 0.413 & 0.02 (-0.03, 0.08) \\
\quad \$100{,}000 or more
& 0.33 (0.13) & 0.010** & 0.07 (0.02, 0.12) \\
\midrule

Age group \\
\quad 18--29
& -- & -- & -- \\
\quad 30--44
& -0.45 (0.13) & <.001*** & -0.09 (-0.13, -0.04) \\
\quad 45--59
& -0.67 (0.13) & <.001*** & -0.14 (-0.18, -0.08) \\
\quad 60+
& -1.04 (0.13) & <.001*** & -0.22 (-0.27, -0.17) \\
\midrule

Metro \\
\quad Non-Metro Area 
& -- & -- & -- \\
\quad Metro Area
& 0.08 (0.11) & 0.447 & 0.02 (-0.03, 0.06) \\
\midrule

Gender \\
\quad Male 
& -- & -- & -- \\
\quad Female
& -0.24 (0.08) & 0.003** & -0.05 (-0.09, -0.02) \\
\midrule

Race/ethnicity group \\
\quad White, non-Hispanic
& -- & -- & -- \\
\quad Asian-Pacific Islander, non-Hispanic
& -0.12 (0.14) & 0.386 & -0.03 (-0.09, 0.03) \\
\quad Black, non-Hispanic
& -0.24 (0.10) & 0.022* & -0.05 (-0.10, -0.01) \\
\quad Hispanic
& 0.21 (0.17) & 0.216 & 0.04 (-0.02, 0.11) \\
\quad Other, non-Hispanic
& 0.02 (0.20) & 0.936 & 0.00 (-0.08, 0.09) \\
\bottomrule
\end{tabular}%
}
\end{table}

\subsection{Exposure by Chart Type}

\begin{figure}[tb]
  \centering 
  \includegraphics[width=\columnwidth, alt={A stacked bar chart showing percentage of respondents who said Yes, No, and Unsure when asked whether they had seen each of the chart types.}]{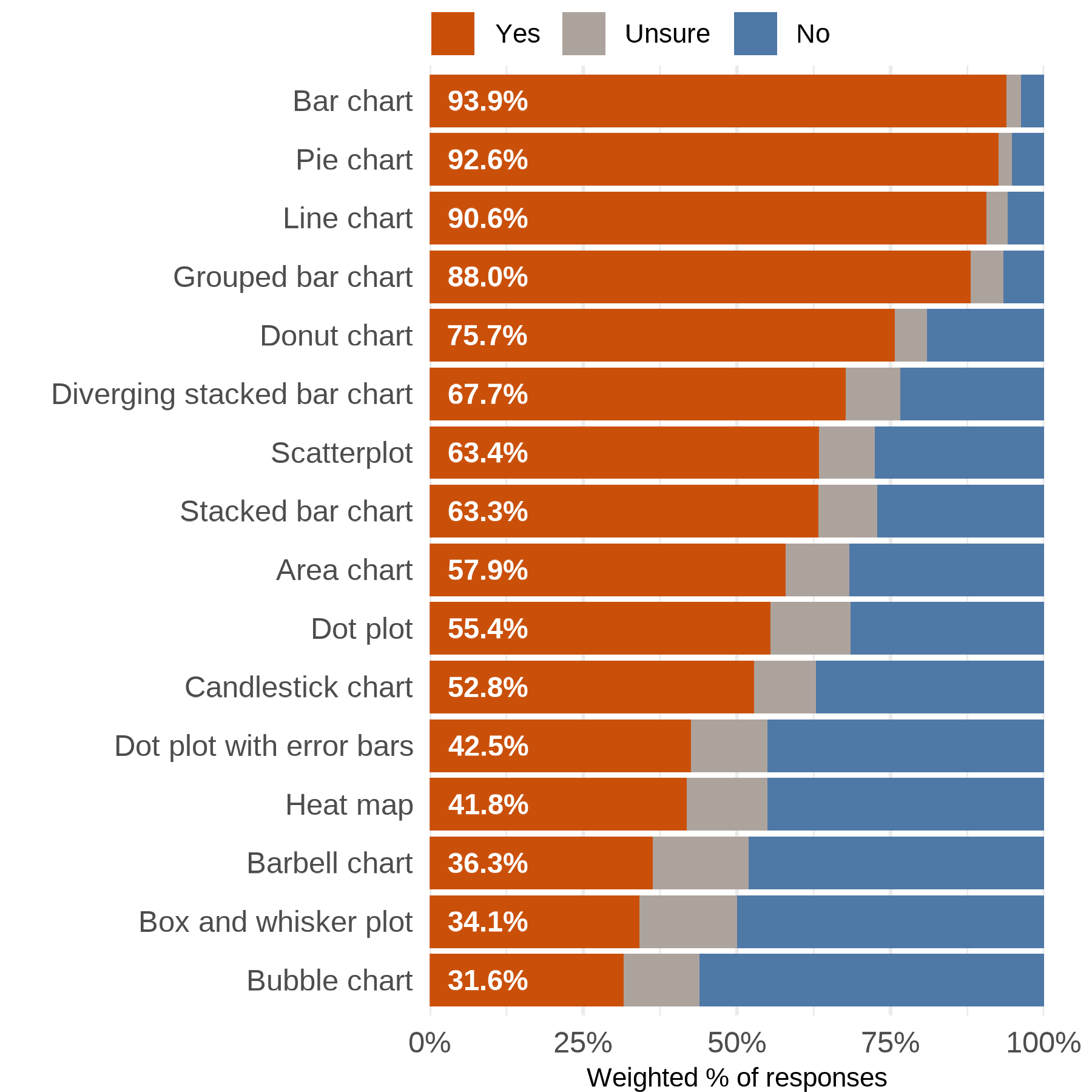}
  \caption{%
  	Visualization of response patterns across sixteen chart types, ordered by highest percentage responding Yes. For each chart type, the weighted percentage of each type of response is shown, with percent Yes labelled in white. %
  }
  \label{fig:overall_familiarity}
\end{figure}

The distribution of Yes, Unsure, and No responses for each specific chart type are shown in \cref{fig:overall_familiarity}. Over 85\% of respondents reported exposure to the four most well-known chart types: bar, pie, line, and grouped bar charts. There is a notable drop-off in recognition for the fifth-most familiar, with 75.7\% of respondents reporting having seen a donut chart. Diverging stacked bar charts, scatterplots, stacked bar charts, area charts, dot plots, and candlestick charts were reported as seen by over half of respondents, with percentage of Yes responses falling between 52.8\% and 67.7\%. Less than half of respondents reported having seen the remaining five chart types: dot plots with error bars, heat maps, barbell charts, box and whisker plots, and bubble charts. 

For chart types with lower rates of reported exposure, there are also higher rates of respondents reporting that they are unsure whether they have seen the chart relative to the most common data visualization types. While only 2.3\% and 2.2\% of respondents selected Unsure for bar and pie charts, respectively, over 15\% of respondents reported being unsure whether they had seen barbell charts and box and whisker plots. When chart types are less commonly known, adults not only report having seen them at lower rates, but are less sure about whether they have seen them. This pattern may reflect limited differentiation among similar chart forms (for example, difficulty distinguishing between barbell charts and dot plots with error bars) or an overall lack of name recognition for these chart types.

\subsection{Chart Exposure by Demographic Group}

\begin{figure}[tb]
  \centering 
  \includegraphics[width=\columnwidth, alt={A connected line chart showing the percentage of respondents who are familiar with each chart type, with a ribbon denoting plus or minus one standard error around the estimate.}]{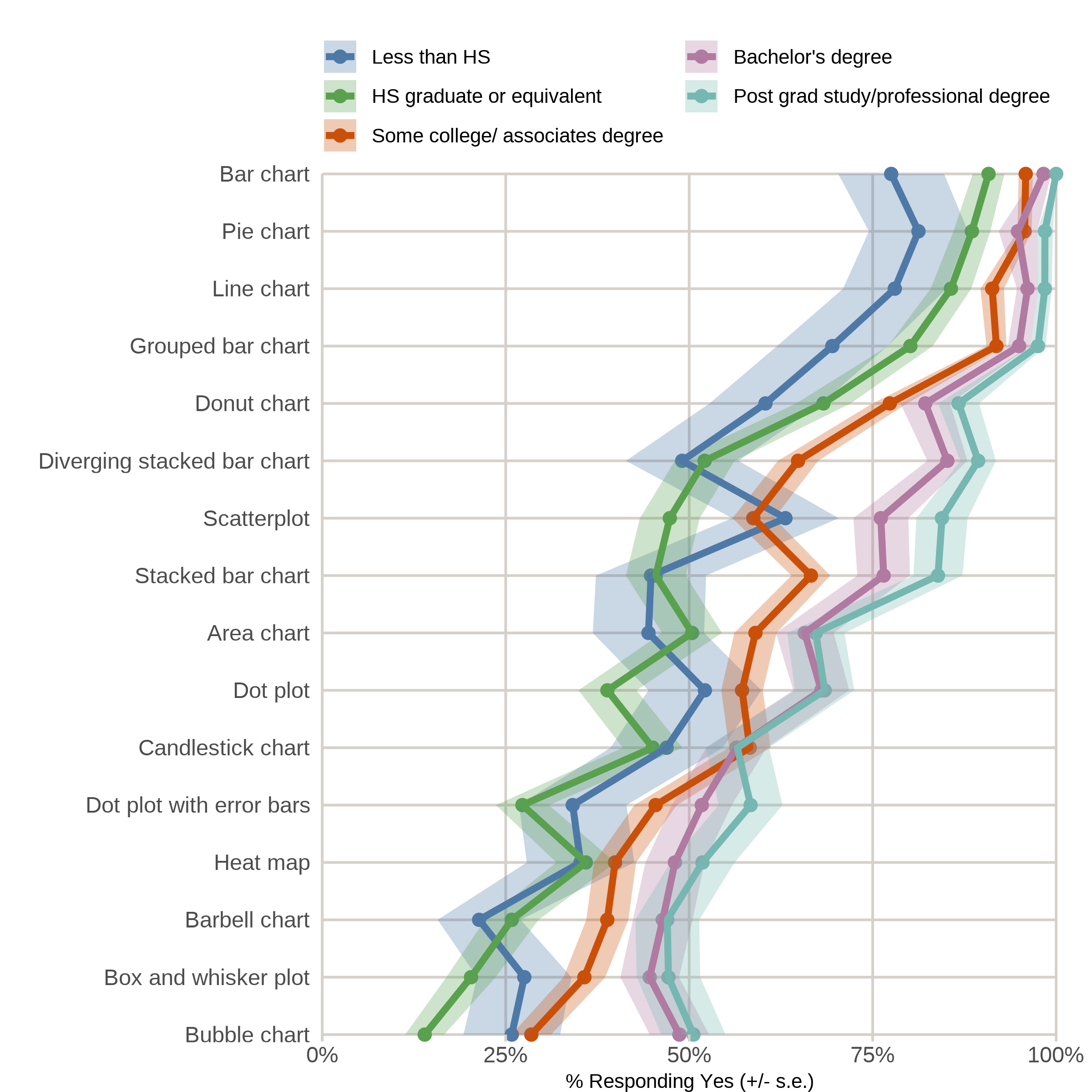}
  \caption{%
  	Weighted percent of respondents (plus or minus one standard error) reporting exposure to each chart type by highest level of educational attainment. %
  }
  \label{fig:familiarity_educ}
  
\end{figure}

Given the pronounced differences in overall exposure to data graphics by education level, \cref{fig:familiarity_educ} presents exposure to each chart type by educational attainment. The results indicate that, across all chart types, more highly educated individuals are more likely to report exposure to a given graphic. Several nuances, however, are worth noting. First, for the four most commonly recognized chart types (bar charts, pie charts, line charts, and grouped bar charts), respondents with a high school diploma or equivalent are more likely to report exposure to these graphics than those with less than a high school education. It is also worth noting that for these commonly used graphics, exposure among respondents with any college coursework is very high and similar to the exposure for respondents with a bachelor’s degree or above. These patterns could suggest that these chart types are more widely used in high school curricula, whereas the less common chart types appear less frequently.

For rarer chart types, a clearer distinction emerges between individuals with at least a bachelor’s degree and those without one. In many cases, respondents with some college experience or an associate’s degree report higher levels of exposure to these charts than those with only a high school education or less, but these differences are generally smaller than the divide between individuals with and without a bachelor’s degree. In several instances, there is no difference between individuals with a high school education or less and those with some college.

Across most chart types, reported exposure among respondents with a bachelor’s degree is indistinguishable from that of respondents with postgraduate education. Exceptions include diverging stacked bar charts, scatterplots, and stacked bar charts, for which postgraduate respondents exhibit higher exposure. This pattern may suggest that these types of charts are more commonly encountered in graduate level coursework.

Next, we fit item-specific logistic regression models to understand how demographic characteristics impact the probability of a participant responding 'Yes'. For each chart type, we create a binary variable encoding (Yes responses compared to No or Unsure responses). For chart type $i$ and respondent $k$, we model: 

\begin{equation}
\label{eq:logistic_model_chart_type}
\mbox{logit}\ P(Y_{ik} = 1) = \mu_i + X_k\mathbf{\beta}_i
\end{equation}

where $X_k$ is the vector of demographic characteristics for respondent $k$. Chart type is denoted by $i = \{ 1, 2, \dots, 16 \}$ where 1 = bar chart, 2 = pie chart, and so forth. Each model is fit separately for each chart type, with the bar chart model fit as: 

\begin{equation}
\mbox{logit}\ P(Y_{1k} = 1) = \mu_1 + X_k\mathbf{\beta}_1
\end{equation}

\begin{figure*}[tb]
  \centering 
  \includegraphics[width=\textwidth, alt={A visualization of average marginal effects from the chart-level logistic regression models estimating the log odds of reporting exposure to each chart type. Positive average marginal effects are shown in darker green, representing the average estimated change in the probability of responding 'Yes' to having seen a chart, while negative average marginal effects are shown in darker orange and represent the average estimated decrease in the probability. Statistically significant effects have a black border around them.}]{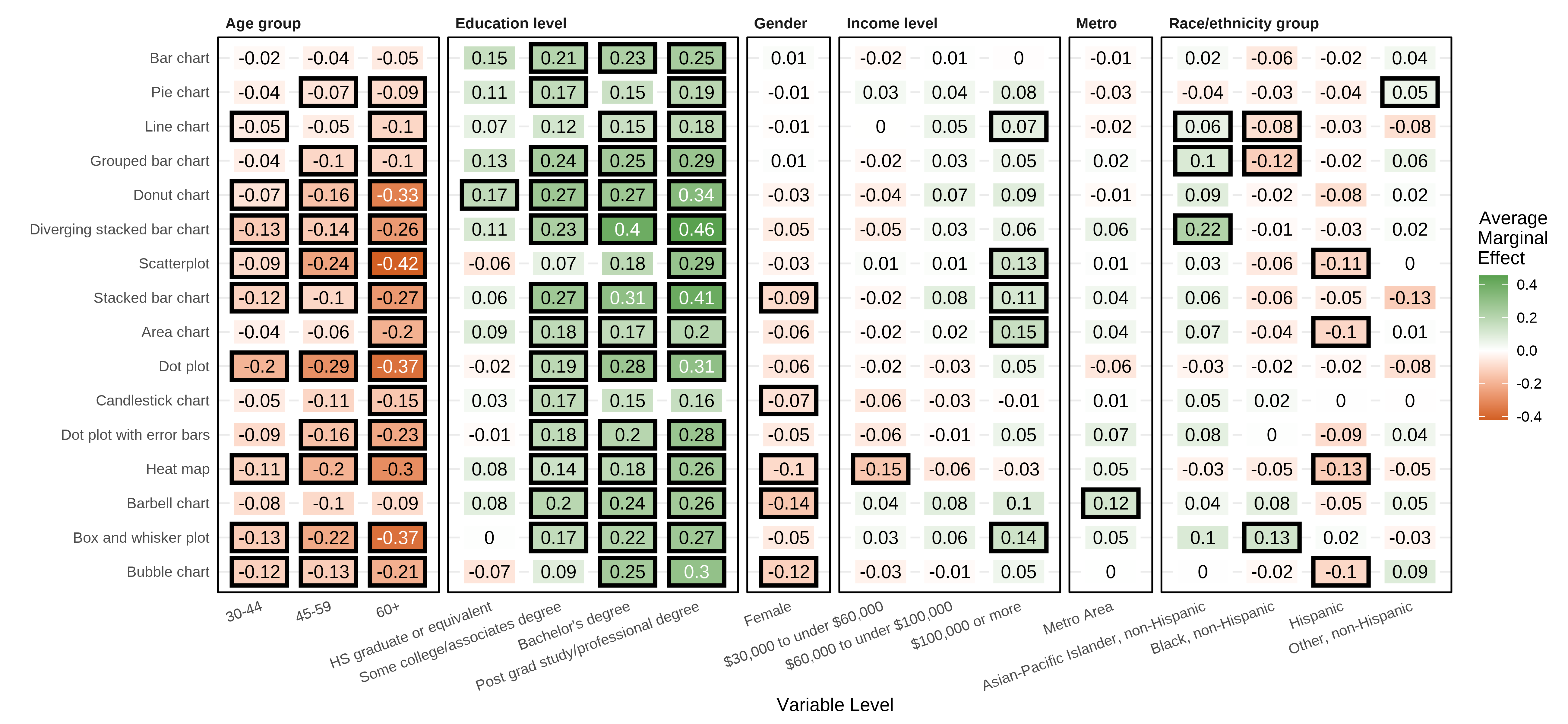}
  \caption{%
  	Average marginal effects from the chart-level logistic regression models estimating the log odds of reporting exposure to each chart type, as defined in \cref{eq:logistic_model_chart_type}. Average marginal effects shown represent the average expected change in the probability of responding 'Yes' on a chart type for the corresponding demographic category, as compared to the reference group. Effects that significantly differ from zero are denoted with a black border. %
  }
  \label{fig:full_model_results}
  
\end{figure*}

Resulting average marginal effects across the sixteen chart type models are visualized in \cref{fig:full_model_results}. Across most chart types, an increase in level of education is associated with a positive increase in the probability of responding 'Yes', though the high school graduate or equivalent group does not have significantly different outcomes than the less than high school group for fifteen out of sixteen chart types. After accounting for education level, income level is generally not associated with a change in the probability of reporting exposure, with exceptions for specific income levels in six chart types. 

Across almost all chart types, older respondents are less likely to report exposure to the chart. The two exceptions are bar charts (which have generally very high exposure across all respondents) and barbell charts. Moreover, for most chart types, exposure declines progressively with age and each older age group reports lower exposure than the one before it. However, for some chart types such as line charts and candlestick charts there are no significant differences across age groups until the 60+ category, which reports the lowest exposure for all chart types. 

Metropolitan area residency, gender and race and ethnicity do not demonstrate systematic significant differences in reported exposure after accounting for education, income, and age group; however, there are specific chart types for which significant differences are observed for specific groups. Notably, women are slightly less likely to report exposure to five chart types than men: stacked bar charts, candlestick charts, heat maps, barbell charts, and bubble charts. Hispanic adults are slightly less likely to report exposure to four types (scatterplots, area charts, heat maps, and bubble charts), Asian-Pacific Islander adults are more likely to report exposure to three common types (line charts, grouped bar charts, and diverging stacked bar charts), and non-Hispanic Black adults are less likely to have seen two of the most common (line charts and grouped bar charts), but more likely to have seen box and whisker plots.

\subsection{Relationships Between Chart Types}

\begin{figure}[tb]
  \centering 
  \includegraphics[width=\columnwidth, alt={Conditional probability matrix.}]{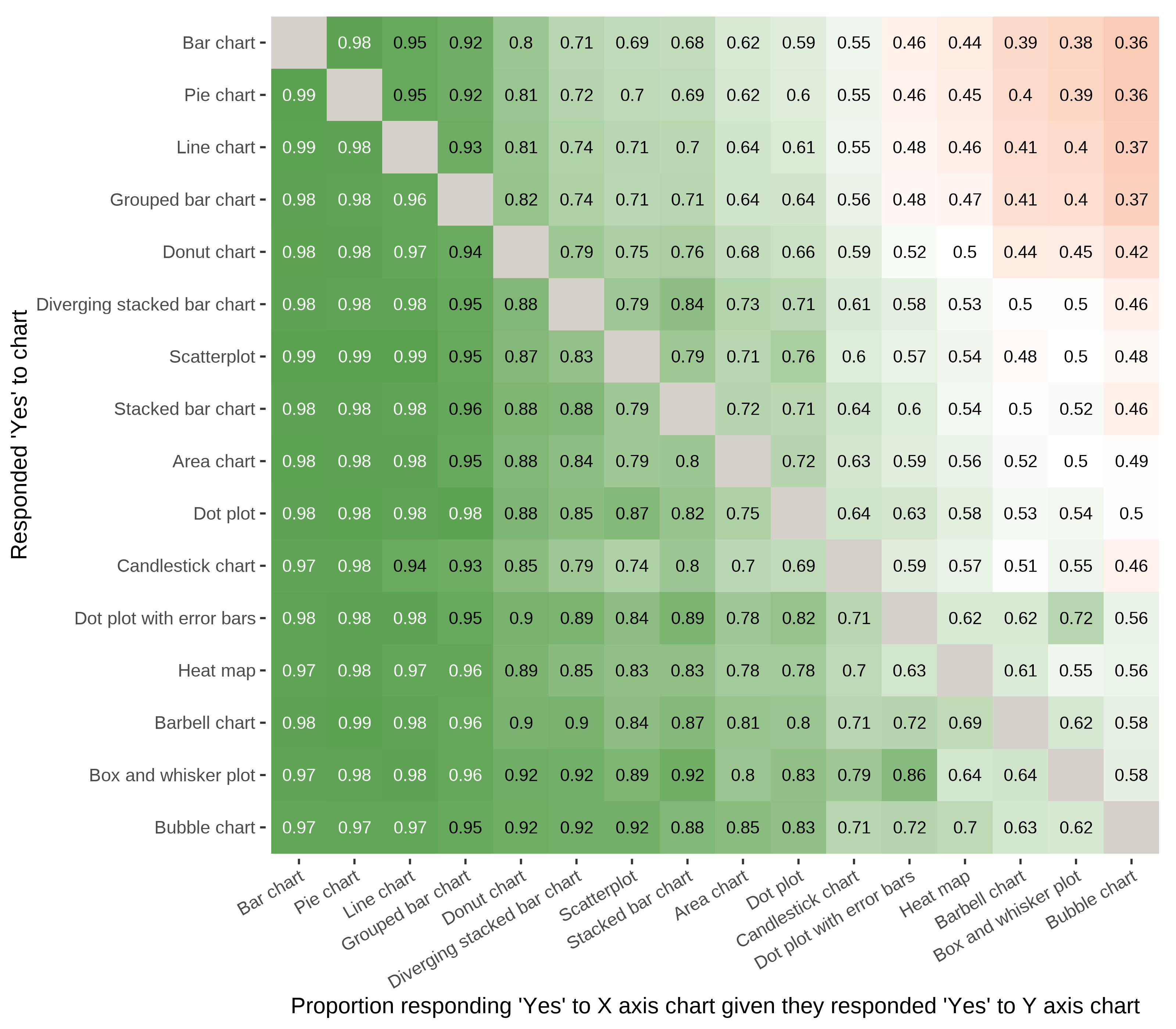}
  \caption{%
  	Conditional probability matrix showing the probability of a respondent knowing the chart shown on the x axis, given that they said they knew the chart shown on the y axis. %
  }
  \label{fig:conditional_probs}
\end{figure}

In addition to measuring chart-specific exposure, it is also important to understand how exposure relates across specific chart types; do certain groupings or pairings of charts have similar exposure? To investigate this question, \cref{fig:conditional_probs} plots the proportion of respondents reporting exposure to the chart type shown on the x axis conditional on the respondent reporting exposure to the chart type shown on the y axis. For example, the bottom left cell of \cref{fig:conditional_probs} indicates that out of respondents exposed to bubble charts, 97\% have been exposed to bar charts.

Unsurprisingly, exposure to the four most common chart types (bar, pie, line, and grouped bar charts) are highly correlated. For example, 99\% of individuals reporting exposure to pie charts also report exposure to bar charts, 98\% of respondents reporting exposure to line charts also report exposure to pie charts, and 96\% of individuals reporting exposure to grouped bar charts report exposure to line charts. Outside of the relationship among the top four most selected chart types, each of the other chart types exhibits a weaker relationship; that is, there is not a strong relationship between recognition of the top four chart types and the less common chart types. 

However, there are some examples of moderate to strong relationships among the less common chart types which are suggestive that exposure to certain chart types may happen in conjunction. For example, of respondents reporting exposure to box and whisker plots, 86\% also report exposure to a dot plot with error bars, even though only 42.5\% of individuals overall reported reported exposure to dot plots with error bars. In addition, of respondents reporting exposure to stacked bar charts, 88\% report exposure to diverging stacked bar charts. Scatterplots are also predictive of exposure with multiple chart types. Of respondents reporting exposure to scatterplots, 87\% report exposure to donut charts, 85\% report exposure to diverging stacked bar charts, 79\% report exposure to stacked bar charts, and 76\% report exposure to dot plots.

Also of note, the candlestick chart shares relatively less of a relationship with other items when compared to other chart types, suggesting that the population's exposure to candlestick charts may occur differently than their exposure to other chart types.  

\section{Discussion}

Overall, there is relatively high baseline exposure to common data visualization types among U.S. adults. The median respondent reported exposure to 10 out of the 16 chart types, and the vast majority (94\%) of participants reported having seen four or more types.

However, we find that overall level of exposure differs significantly across demographic groups. Higher levels of education are associated with exposure to more chart types, reflecting increased exposure to data visualization through formal educational pathways. Older age groups report exposure to fewer chart types on average, potentially reflecting the increase in data visualization as a communication method in the last several decades. Older adults may also have been less likely to experience a broader set of data visualizations during primary, secondary, and post-secondary education as they were used less widely at the time. While education and age are the biggest drivers of exposure, there are also differences across many population subgroups. Women and Black non-Hispanic adults both report slightly lower exposure overall, while people in high-income households (\$100,000 or more) report slightly higher exposure overall. These patterns represent a key finding for visualization literacy: literacy may vary greatly in the population partially because different groups have distinct levels of exposure to data visualization in the first place, regardless of their skill level at reading or interpreting quantitative information. While less exposure does not inherently imply lower literacy, it is likely that literacy improves as individuals become more familiar with a given mode of visualization. 

We also uncover stark differences in exposure across individual chart types. Unsurprisingly, bar charts, pie charts, and line charts have a high level of exposure among U.S. adults. However, other chart types suffer from a lack of permeation among the population; less than half of adults have seen five of the chart types we tested (dot plots with error bars, heat maps, barbell charts, box and whisker plots, and bubble charts). These differential levels of exposure highlight a key consideration for understanding visualization literacy: if overall literacy is measured using primarily the most common chart types, that measure of literacy may not extend to data visualization \emph{in general}. It is therefore essential for practitioners to consider the exposure of their audience to a given chart type when designing and implementing data visualizations.  

Exposure to specific charts is further fractured across demographic groups; not only are there differences in exposure across demographic groups, but those exposure patterns also differ across chart types! While there are sweeping patterns of differing exposure by education and age, those differences are not universal. Both bar charts and barbell charts do not differ significantly across age groups, likely for conflicting reasons: bar charts are generally familiar to all age groups as the most commonly-recognized chart type, while barbell charts are uncommon across all age groups, and younger adults do not share in any benefit of greater exposure to barbell charts. While women had slightly lower recognition overall on the sixteen items, we see that it is isolated to only small differences in exposure to five out of 16 types. Idiosyncratic differences appear across race and ethnicity groups as well, with non-Hispanic Black adults showing slightly lower exposure on two common types but slightly higher exposure on a rarer type, Hispanic adults showing slightly lower exposure with four specific types, and Asian-Pacific Islander adults showing higher exposure on three relatively common types, as compared to non-Hispanic White adults. High-income households do have slightly higher exposure to five different types; however, metropolitan area residency only benefits barbell charts. These complex relationships lay the groundwork for approaching visualization literacy gaps; with this clearer understanding of who has low exposure to specific chart types, we can better target both efforts to increase visualization literacy as well as better design visuals for specific audiences. For example, our results show very low exposure to scatterplots, dot plots, and box and whisker plots among older adults, and would suggest that those types should not be used as a primary form of visual communication among that group unless additional training is provided. Similarly, scatterplots and bubble charts both have substantial increases in exposure among those with Bachelor's degrees or higher; it may be beneficial to consider incorporating those chart types earlier in educational curricula so as to train a broader set of adults in their use.   

The fact that exposure gaps between demographic groups vary by chart type implies that visualization literacy gaps across groups may be even more pronounced for lesser-known chart types than for the most familiar types. Since prior studies have demonstrated gaps in visualization literacy across education groups for bar charts, pie charts, and line charts, our results signal a substantial concern for visualization literacy in general -- for less widely-known chart types, literacy gaps may be even larger due to lack of familiarity and comfort with chart types.

We also see that exposure to some chart types may be linked; while exposure to common types (bars, pies, lines) is not shown to impact exposure to rarer types, exposure to rarer types is indicative of higher overall exposure across other chart types. We also observe some specific groupings of exposure to charts, such as the strong co-occurrence of box and and whisker plots and dot plots with error bars. While this is not particularly surprising, it demonstrates that visualization exposure is not a monolithic concept. Outside of the common chart types, drop-offs in exposure separate out those who have broader exposure to data visualization generally, those who gain exposure in specific contexts, such as formal educational settings or domain-specific charts, and those who have have been exposed to only the top three or four types.

\subsection{Limitations}

While our study of sixteen chart types covers a substantial portion of charts used in recent proposed visualization assessment tests, it is far from comprehensive of all visual formats used in public data communication. For example, we do not cover maps or more complex network diagrams. Our study also does not have perfect coverage of all recent visualization assessments; we do not include treemaps or icon arrays, two major formats used in some key visualization literacy assessments. 

In addition, there are some limitations in the construct of measured exposure; we show participants a very simple iconographic representation of a chart paired with a name. Participants may not fully understand the chart type or may not recognize the name or icon in relation to something they may have actually seen in the real world, leading to under-measurement of exposure. The increase in reporting of 'Unsure' values for rarer types underscores this limitation; for less common chart types, more adults are simply unsure whether they have seen it just by looking at an icon and the name. Additionally, some chart types go by multiple names or could be conflated with one another. Notably, our iconographic representation of a candlestick chart may be closer to a dynamite plot, and it may be difficult for some people to distinguish between a dot plot with error bars and a box and whisker plot. However, even with these potential measurement difficulties, we still identify substantial and meaningful differences in exposure.

\section{Conclusions and Future Work}

This study establishes population-level estimates of exposure to sixteen common data visualization types among U.S. adults, providing major insights into how the population interacts with data visualizations. Our results demonstrate noteworthy variation both within the population and across visualization types, highlighting that exposure to data visualization is far from uniform and occurs through both formal and informal pathways. 

These findings have important implications for data visualization literacy researchers, educators, and practitioners. Researchers who are working to measure visualization literacy should take into account observed differences in exposure throughout the process -- both when developing tests of literacy and when interpreting their results, as findings on overall visualization literacy may be impacted by underlying patterns of exposure.

Educators could also use these results to reshape how and when data visualizations are introduced to students. The strong relationships between formal education and exposure indicate a lack of exposure in primary and secondary education for many chart types. If some of these less-common chart types were incorporated earlier in educational pathways, this could increase baseline exposure and familiarity across the population.

Practitioners should carefully consider their use of chart types when communicating data with both the general public and specific populations. Because individual groups may have stronger or weaker baseline exposure to specific types, practitioners may want to tailor their use based on their audience, especially when data are used to communicate important information for decision-making. Many less common chart types may be difficult for large portions of the adult population in the U.S. to understand, as they may not have seen them before. While lack of reported exposure to a specific chart type does not inherently imply lack of understanding or interpretive ability with that chart type, unfamiliar charts may take a viewer longer to process or could lead to misunderstanding of the graphical format. For rarer graphics, providing additional explanation or interpretive guidance may be especially important to help mitigate exposure gaps and better support all viewers’ understanding.

There are several directions for future work building upon our findings. First, it would be beneficial to test multiple approaches to measuring exposure. Potential alternatives include showing real-world chart examples for each instead of iconographic images or providing multiple names for chart types that may be ambiguously defined. In addition, asking participants to name the chart rather than providing them the name would provide an understanding of not just whether an individual has seen the chart type, but whether they know what the chart is in a formal sense. And while this study establishes baseline exposure, future work should build a stronger link between exposure, familiarity, and ability by linking visualization literacy assessments and exposure measures. Measuring direct relationships between exposure and understanding will allow the field to most effectively identify where literacy fails and why, including whether a lack of baseline exposure substantially harms literacy.

\acknowledgments{%
	The authors wish to thank Edward Mulrow for his input on the design of this study. 
  This work was supported in part by the National Science Foundation (Grant Nos. 2346660 and 2521777).%
}



\end{document}